\documentclass[preprint,authoryear,5p,11pt]{elsarticle}

\usepackage{amssymb,amsmath,amsfonts,epsfig,color,epstopdf}
\usepackage{algorithm,algorithmic}
\usepackage{xcolor}
\floatname{algorithm}{Algorithm}
\newcommand{\beq}{\begin{equation}}
\newcommand{\eeq}{\end{equation}}
\newcommand{\beqn}{\begin{eqnarray}}
\newcommand{\eeqn}{\end{eqnarray}}

\def\bmath#1{\mbox{\boldmath$#1$}}

\usepackage{natbib}

\usepackage{makecell}
\usepackage{listings}
\usepackage{hyperref}

\usepackage{lineno}

\journal{Astronomy and Computing}

\begin{document}

\begin{frontmatter}



\title{Polarization based direction of arrival estimation using a radio interferometric array}


\author{Sarod Yatawatta}

\affiliation{organization={ASTRON, The Netherlands Institute for Radio Astronomy},
            addressline={Oude hoogeveensedijk}, 
            city={Dwingeloo},
            country={The Netherlands}}

\begin{abstract}
  Direction of arrival (DOA) estimation is mostly performed using specialized arrays  that have carefully designed receiver spacing and layouts to match the operating frequency range. In contrast, radio interferometric arrays are designed to optimally sample the Fourier space data for making high quality images of the sky. 
  Therefore, using existing radio interferometric arrays (with arbitrary geometry and wide frequency variation) for DOA estimation is practically infeasible except by using images made by such interferometers. In this paper, we focus on low cost DOA estimation without  imaging, using a subset of a radio interferometric array, using a fraction of the data collected by the full array, and, enabling early determination of DOAs. The proposed method is suitable for transient and low duty cycle source detection. Moreover, the proposed method is an ideal follow-up step to online radio frequency interference (RFI) mitigation, enabling the early estimation of the DOA of the detected RFI.
\end{abstract}


\end{frontmatter}


\section{Introduction\label{sec:intro}}
Estimating the direction of arrival of incoming radiation plays a major role in array signal processing. In specialized application such as radar, localization, and wireless communications, there are arrays dedicated for DOA estimation. Such dedicated arrays are carefully designed in terms of the receiver placement to get the desired accuracy and sensitivity.

Radio interferometric arrays on the other hand have a different goal, i.e., to observe the celestial sky (or rather its Fourier transform) in order to produce high fidelity images. Therefore, the receiver placement is determined to get a full Fourier space sampling. However, the increased RFI environment that modern radio interferometers encounter creates a need for rapid response RFI mitigation and finding the sources of RFI (or their directions of arrival).

In this paper, we propose a method to use an already existing radio interferometric array as a DOA estimation array without modifying the array layout. The only requirement is to have subsets of dual polarized receivers that are approximately co-linear. This requirement is easily satisfied by phased array receivers where the individual cross-dipole elements are placed on a regular grid. Otherwise, subsets of receivers can be selected if they are approximately co-linear. There is no requirement on the spacing of the receivers (or the baseline length). Most arrays that are designed for DOA estimation have spacing that are less than half a wavelength or sub-arrays that have co-prime spacing. As this spacing changes with the wavelength, their operating frequency range is also limited. In contrast, radio interferometric arrays have a wide frequency range of operation. Therefore, in order to perform DOA estimation with a wide frequency range using a radio interferometric array, we cannot use most of the excising DOA estimation techniques. 

Instead of following a centralized processing approach, we break down the computations required for DOA estimation onto the incoming data flow. At the correlator, where the data streams of two receivers are correlated to produce the visibilities of a baseline, we also perform the first processing step required for the DOA estimation. At this stage, we apply the estimation of signal parameters via rotational invariance \cite[ESPRIT]{ESPRIT} onto the correlated and averaged visibilities. With the use of polarized ESPRIT \citep{Li1991}, we get the sufficient statistic (phase angles) for DOA estimation per each baseline. These angles are collected at a central processing point where phase unwrapping and final DOA estimation using a deep neural network is performed.

The essence of our DOA estimation method and the underlying assumptions can be summarized as follows:
\begin{enumerate}
  \item We apply polarized ESPRIT \citep{Li1991} to data streams of each baseline (or an interferometer). The data of each baseline are formed by correlating data from a pair of receivers that have dual polarized feeds. We assume the incoming signal to have only one dominant source and hence only one DOA. The output of running ESPRIT is the estimate of the phase of the DOA with respect to each baseline.
  \item We combine the phase estimates of the baselines that belong to an (almost) linear sub-array. We use the phase difference projection algorithm \citep{Hui2019,Hui2021} to unwrap the phase estimates of each sub-array. We assume the full-array can be broken down into several (almost) linear sub-arrays and apply the phase unwrapping to each sub-array.
  \item Finally, we combine the unwrapped phase of all sub-arrays to jointly estimate the 2D DOA of the dominant source. This is formulated as a minimization of a non-linear cost function. However, this cost function is non-convex with several local minima. In order to find a refined estimate, we train a transformer deep neural network (DNN) \citep{Vaswani}. 
\end{enumerate}
The relation of the proposed method to prior work can be listed as follows:
\begin{itemize}
  \item There are already specialized methods for DOA estimation in radio interferometry. Most of them  such as spatial filtering, beamforming, or matrix factorization \citep{raza2002spatial,boonstra2005spatial,BrossardDOA,Steeb2018,ADrosz2021} require centralized processing (assembling a full correlation matrix) but our method is far less computationally demanding. It is also possible to consider DOA estimation as a source fitting \citep{Bruce2022} or a specialized image synthesis \citep{Ducharme2025} that require the accumulation of sufficient data to perform an image synthesis. In contrast, our method is inherently distributed (i.e., baseline based), and is suitable for online operation with only a fraction of the data received by the array (i.e., not all baselines are needed). This also becomes an ideal follow-up of online RFI detection \citep{Y2024} where we can only use the data corrupted by RFI as the input to the DOA estimation. However, our method requires the interferometric array to have dual polarized receivers and also a sufficient number of receivers that are approximately co-linear. As a limitation, our method is only capable of detection of a single DOA, which we consider to be the dominant source.
  \item Most arrays that are dedicated for DOA estimation are designed with a specialized layout of receivers, such as with uniform and linear, circular or rectangular receiver spacing, or with sub-arrays that have receivers with co-prime spacing. These specialized layouts substantially reduce the computational cost of DOA estimation enabling the use of methods like multiple signal classification \citep[MUSIC]{MUSIC}. On the other hand, with arbitrary array geometries (or arrays that are not dedicated to DOA estimation), DOA estimation is more expensive and popular methods exploit sparsity \citep{YANG2018,Yimin2024}, or perform maximum likelihood estimation \citep{CHUNG2002}. Moreover, rather than specializing the receiver layout, the receiver capabilities can be enhanced such as by using 3D sensors \citep{wen2025fast} or phase interferometers \citep{DING2026} for DOA estimation with arbitrary arrays. In contrast, our method is capable of low cost DOA estimation using existing arrays dedicated for radio astronomy.
  \item Most specialized DOA estimation arrays have their receivers spaced by half a wavelength ($\lambda/2$), and sub-arrays that have co-prime spacing. In wideband operation, the receiver spacing changes with frequency, needing phase unwrapping \citep{Hui2019,Hui2021}. Radio astronomical arrays are inherently wideband, and in order to use radio astronomical arrays for wideband DOA estimation, we also need phase unwrapping. We note that the wideband data are channelized into a set of discrete frequencies at the correlator and we consider the DOA estimation for each channel individually (assuming a narrow band source). For each channel, we perform phase unwrapping using the phase difference projection algorithm \citep{Hui2019,Hui2021} that is done as a pre-computation (i.e., the phase unwrapping computations are only dependent on the array configuration and are independent of the data). In order to perform phase unwrapping, we select sub-arrays from the available receivers that are approximately co-linear. This ability to create sub-arrays that are almost co-linear is an additional requirement for our method.
  \item Combining classic signal processing (model driven approaches) with deep learning (data driven approaches) is an emerging area of research \citep{DAMUSIC,Wang2024,Bell2025}. The proposed method improves on existing similar methods by working with arbitrary array layouts and with the use of a transformer DNN. A demanding requirement of such methods is the need for training but on the positive side, a properly trained DNN is capable of accumulating knowledge of multiple observations to make an informed decision in contrast to a purely model based approach that relies exclusively on the current observation.
\end{itemize}

The rest of the paper is organized as follows. In section \ref{sec:model}, we provide details of the signal processing model for DOA estimation. Next in section \ref{sec:DNN}, we describe the transformer deep neural network that is used in the last step of DOA estimation. In section \ref{sec:simul}, we provide results based on simulations using two existing radio interferometric arrays, namely, \citep[AARTFAAC]{AARTFAAC} and \citep[SKA-Low]{SKA0}. Finally, in section \ref{sec:conc}, we draw our conclusions.
 
{\em Notation}: Lower case bold letters refer to column vectors (e.g. ${\bmath y}$). Upper case bold letters refer to matrices (e.g. ${\bf {C}}$). The sets of integer, real and complex numbers are denoted by $\mathbb{Z}$, $\mathbb{R}$ and $\mathbb{C}$, respectively.  The matrix transpose, and Hermitian transpose are referred to as $(.)^{T}$ and $(.)^{H}$, respectively. The statistical expectation operator is given as $E\{.\}$. The $i$-th to $(j-1)$-th elements of a vector ${\bmath y}$ is given by ${\bmath y}[i:j]$. The $l_2$ norm is given by $\|.\|$. A uniform distribution in $[0,1]$ is given by $\mathcal{U}(0,1)$. A Gaussian distribution with mean $\mu$ and standard deviation $\sigma$ is given by $\mathcal{N}(\mu,\sigma)$.

\section{The data model\label{sec:model}}
In this section, we provide a brief overview of the signal processing model based on 
\cite{Ebadi2024,Molaei2024}.

\begin{figure}[ht]
  \begin{minipage}{0.98\linewidth}
    \begin{center}
      \epsfig{figure=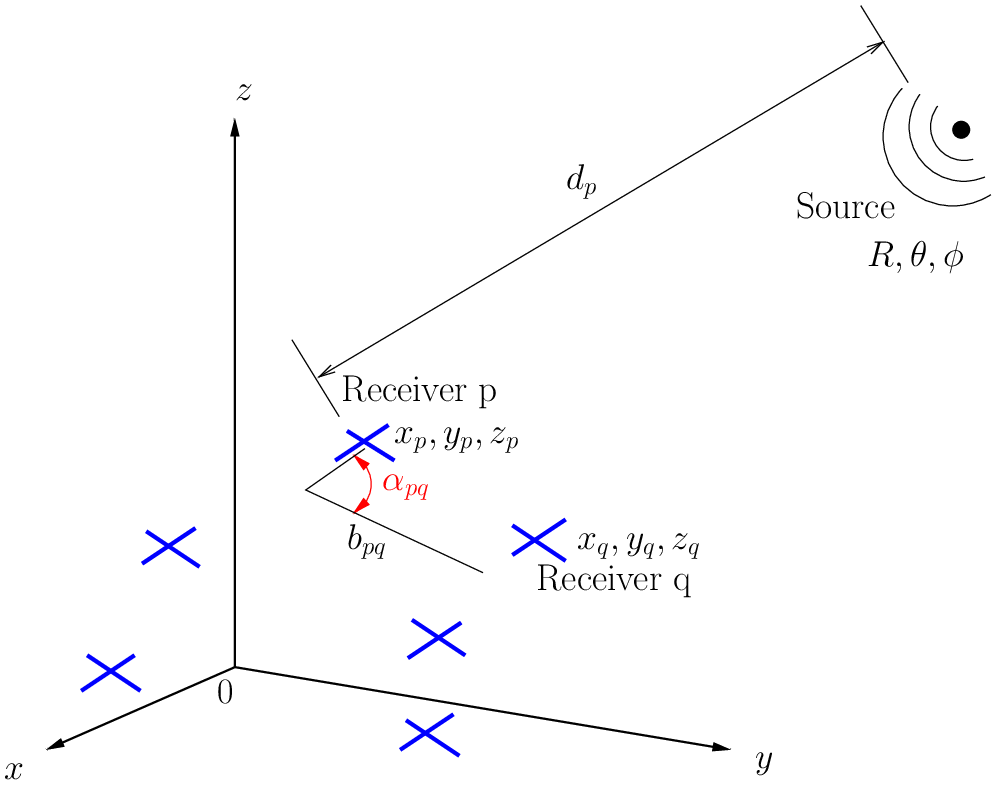,width=8.0cm}\\
    \end{center}
  \end{minipage}
  \caption{Radiation from a source at distance $R$ and angle $\theta,\phi$ is received by a 3D array of arbitrarily placed dual polarized receivers (blue crosses \textcolor{blue}{x}). The coordinates of the $p$-th receiver are given by ${\bf x}_p$. The baseline $pq$ is formed by correlating the digitized and channelized data received by receivers $p$ and $q$. The distance to the source from the $p$-th receiver is given by $d_p$. The baseline length is $b_{pq}$. The angle between the baseline $pq$ and the source DOA is $\alpha_{pq}$.\label{fig:geom}}
\end{figure}

We consider a source at range $R$ from the origin whose signal is received by an array of receivers that are arbitrarily located in 3D space as shown in Fig. \ref{fig:geom}. The unit vector describing the source direction $\widehat{\bf s}$ ($\in \mathbb{R}^3$) can be given as
\beq \label{s}
\widehat{\bf s}=\left[ \begin{array}{c} 
\cos\theta \cos\phi\\
\cos\theta \sin\phi\\  
\sin\theta \end{array} \right]
\eeq
where the angles $\theta$ and $\phi$ describe the source direction in elevation and azimuth.
The dual polarized receiver $p$ is at position ${\bf x}_p$ ($\in \mathbb{R}^3$) which is given by
\beq \label{x}
{\bf x}_p=\left[ \begin{array}{c}
x_p\\
y_p\\
z_p
\end{array} \right].
\eeq
The distance to the source from the $p$-th receiver is given by
\beq \label{d}
d_{p}=\|{\bf x}_p - R\  \widehat{\bf s}\|
\eeq
and the delay of the signal received at the $p$-th receiver is
\beq
\tau_{p} = \frac{d_{p}}{c}
\eeq
where $c$ is the speed of light (free space propagation).

The array factor for receiver $p$ is given by \citep{Ebadi2024}
\beq
a_p(\theta,\phi)=\frac{d_p}{R}\exp({\jmath 2\pi f \tau_p})\ \in \mathbb{C}
\eeq
where $f$ is the observing frequency. We consider data that are channelized into narrow band channels.
The intrinsic polarization of the source is given by  $\gamma$ and $\eta$ ($\in \mathbb{R}$). The receiver beam sensitivity and the rotation (from reference polarization orientation) is described by ${\bf E}_{p}$ ($\in \mathbb{C}^{2\times 2}$).

The channelized instantaneous voltage at receiver $p$ for the two polarizations, ${\bf v}_p(t) \in \mathbb{C}^{2}$, is given by
\beq \label{voltage}
{\bf v}_p(t)=a_p(\theta,\phi) E_s {\bf E}_{p}\left[ \begin{array}{c}
  -\cos\gamma\\
  \sin\gamma\  e^{ \jmath \eta}
\end{array}
\right]
e^{\jmath (2 \pi f t + \psi_0)} + {\bf n}_p(t)
\eeq
where $E_s$ is the intrinsic intensity of the source and $\psi_0$ is the arbitrary phase (set to $0$ for simplicity). The noise is given by ${\bf n}_p(t)$ ($\in \mathbb{C}^2$) that we assume to have a zero mean, complex circular Gaussian distribution with variance $\sigma^2$. We define the signal to noise ratio (SNR) of the received signal as $E_s\|{\bf E}_p\|/\sigma$.

The correlation of the signals received at receivers $p$ and $q$ ($p\ne q$) is formed as follows. We combine the discretized voltages into a vector ${\bf v}_{pq}$ ($\in \mathbb{C}^4$) as
\beq
{\bf v}_{pq}=\left[ {\bf v}_p^T,\ {\bf v}_q^T \right]^T
\eeq
and its correlation ${\bf R}_{pq}$ ($\in \mathbb{C}^{4 \times 4}$) is formed as
\beq
{\bf R}_{pq}=E\{ {\bf v}_{pq} {\bf v}_{pq}^H \} \approx \frac{1}{T}\sum {\bf v}_{pq} {\bf v}_{pq}^H 
\eeq
where $T$ samples are averaged to estimate ${\bf R}_{pq}$.

The eigenvalue decomposition of ${\bf R}_{pq}$ can be expressed as \citep{Li1991}
\beq \label{eig}
{\bf R}_{pq} {\bf V}_{pq}={\bf V}_{pq}{\bmath{\Lambda}}_{pq}
\eeq
where the eigenvalues are in the diagonal of ${\bmath{\Lambda}}_{pq}$ diagonal matrix, and the eigenvectors form the columns of ${\bf V}_{pq}$.
Selecting the dominant eigenvalue and its eigenvector as ${\bf v}$ ($\in \mathbb{C}^4$), we create vectors
\beq 
{\bf e}_x={\bf v}[0:2],\ \ {\bf e}_y={\bf v}[2:4]\ \in \mathbb{C}^{2}
\eeq
and extract the phase $\widehat{\psi}_{pq}$ as 
\beq \label{psi}
\widehat{\psi}_{pq}=\angle\left(\frac{{\bf e}_x^H {\bf e}_y}{{\bf e}_x^H {\bf e}_x}\right).
\eeq
This is the essence of the polarized ESPRIT algorithm \citep{Li1991} and note that we do not need information about the source parameters ($\gamma$,$\eta$,$E_s$) or the beam polarization (${\bf E}_p$,${\bf E}_q$).

We assume the source to be in the far field of baseline $pq$, i.e., $d_{p}, d_{q} \gg \frac{2 b_{pq}^2} {\lambda}$ \citep[Fraunhofer limit]{Guerra2021}. Note that for the full array this limit might be larger, but here it is the baseline length $b_{pq}$ that determines the Fraunhofer limit and is smaller than the full array.
With this assumption, the estimated phase $\psi_{pq}$ can be approximated as 
\beq \label{wrapped_psi}
\psi_{pq}\approx 2 \pi \frac{b_{pq}}{\lambda} \cos \alpha_{pq} + 2k_{pq} \pi
\eeq
where $\alpha_{pq}$ is the angle formed by the source DOA $\widehat{\bf s}$ and the baseline unit vector $\widehat{\bf b}_{pq}$. For baseline lengths larger than $\lambda/2$, which is almost always the case as we have no control over the baseline selection, the true phase can be larger than $2\pi$ hence we estimate the wrapped phase. The last term in (\ref{wrapped_psi}) denotes phase wrapping, where $k_{pq}$ ($\in \mathbb{Z}$) is an unknown integer. In order the unwrap the phase estimate, we use the phase difference projection algorithm \citep[PDP]{Hui2019,Hui2021}. The requirement for applying this algorithm is having a set of receivers that are almost co-linear. Depending on  the original array, we can find (almost) co-linear sub-arrays by selection. For an array with receivers that are already on a regular grid, this task is simple. For an array with randomized receiver positions, we can apply algorithms such as random sample consensus \citep[RANSAC]{RANSAC} for this purpose. We provide examples illustrating this in section \ref{sec:simul}.
 
With the use of phase different projection to the almost co-linear sub-arrays, we get the unwrapped phase  $\phi_{pq}$ from the wrapped phase $\psi_{pq}$ as 
\beq \label{phi}
\phi_{pq}=\psi_{pq}+2\widehat{k}_{pq}\pi
\eeq
where $\widehat{k}_{pq}$ is calculated by the  PDP algorithm for each sub-array. Note that the bulk of the computations of the PDP algorithm is independent of the data and is pre-calculated as explained in \cite{Hui2021}.

With the unwrapped phase estimate of (\ref{phi}), we can combine the phase estimates of all baselines to find the DOA, i.e., $\theta,\phi$. In order to do this, we create the combined cost function $f(\theta,\phi)$,
\beq \label{fcost}
f(\theta,\phi)=\sum_{pq} \left( \left(\widehat{\bf b}^T_{pq}\ \widehat{\bf s} \right)^2 - \left(\frac{\phi_{pq}}{2 \pi b_{pq}/\lambda}\right)^2 \right)^2 
\eeq
and the global minimum of this should give us the correct estimate for $\theta$ and $\phi$. Unfortunately, $f(\theta,\phi)$ has multiple local minima and a straight forward optimization strategy to find $\theta$ and $\phi$ does not work. On top of this, the noise, and phase unwrapping errors could also lead to an incorrect solution. Therefore, to find the best estimates for $\theta$ and $\phi$, we follow a data driven approach where we train a deep neural network for this task as described in section \ref{sec:DNN}.

Furthermore, it is also possible to formulate the cost function (\ref{fcost}) to include the range $R$ as well as $f(R,\theta,\phi)$, and using this formulation to estimate the range in addition to the DOA for near-field sources is discussed in \ref{appendixA}.
\section{Transformer deep neural network\label{sec:DNN}}
The transformer DNN architecture \citep{Vaswani} has expanded into various deep learning applications since its introduction for natural language processing. 
Recent examples of the use of transformer in array signal processing \citep{Lan2022,Lan2023,LIU2023,Wang2024,wu2024two,Fu2024,zhao2025} also include applications of DOA estimation, albeit with specialized array geometries dedicated for DOA estimation. The proposed method differs from such previous applications of the transformer architecture in dealing with arbitrary array shapes (in 3D) and the use of the transformer only as an enhancement to the model based signal processing as described in section \ref{sec:model}.

We give a signal processing oriented description of the transformer architecture following \citep{Kim2023}. The basic building block of the transformer architecture is the multi-head attention mechanism. We consider the input to the multi-head attention to be three matrices, i.e., the query ${\bf X}_{\mathrm{q}}$ ($\in \mathbb{R}^{N_q\times d_{\mathrm{model}}}$), the key ${\bf X}_{\mathrm{k}}$ ($\in \mathbb{R}^{N_{kv}\times d_{\mathrm{model}}}$), and the value ${\bf X}_{\mathrm{v}}$ ($\in \mathbb{R}^{N_{kv}\times d_{\mathrm{model}}}$) and its output to be a matrix ${\bf X}_{\mathrm{out}}$ ($\in \mathbb{R}^{N_q\times d_{\mathrm{model}}}$). The inner dimension $d_{model}$ (also called the embedding dimension) determines the learnable parameters within the multi-head attention block. The embedding dimension is further divided into $N_{\mathrm{head}}$ equal subsets of size $d_{\mathrm {head}}$, therefore $d_{\mathrm {head}}=d_{model}/N_{\mathrm{head}}$. Note that $d_{model}$ should be a multiple of $N_{\mathrm{head}}$.

For each head, the query, key and value matrices are subdivided. For the $n$-th head, we have the query, the key and the value matrices as ${\bf X}_{qn}$ $\in \mathbb{R}^{N_q\times d_{\mathrm{head}}}$, ${\bf X}_{kn}$ $\in \mathbb{R}^{N_{kv}\times d_{\mathrm{head}}}$, and ${\bf X}_{vn}$ $\in \mathbb{R}^{N_{kv}\times d_{\mathrm{head}}}$, respectively. The output of the $n$-th block is formed by first forming the query, key, value triplet as
\beqn
{\bf Q}_n={\bf X}_{qn}{\bf W}_{qn} \in \mathbb{R}^{N_q\times d_{\mathrm{head}}}\\\nonumber
{\bf K}_n={\bf X}_{kn}{\bf W}_{kn} \in \mathbb{R}^{N_{kv}\times d_{\mathrm{head}}}\\\nonumber
{\bf V}_n={\bf X}_{vn}{\bf W}_{vn} \in \mathbb{R}^{N_{kv}\times d_{\mathrm{head}}}
\eeqn
where the learnable parameters are ${\bf W}_{qn}, {\bf W}_{kn}, {\bf W}_{vn} \in \mathbb{R}^{d_{\mathrm{head}} \times d_{\mathrm{head}}}$. After forming ${\bf Q}_n$, ${\bf K}_n$ and ${\bf V}_n$, the output is calculated as
\beq
{\bf X}_n=\mathrm{softmax}\left(\frac{{\bf Q}_n {\bf K}_n^T}{\sqrt{d_{head}}}\right) {\bf V}_n \in \mathbb{R}^{N_q\times d_{\mathrm{head}}}
\eeq
where $\mathrm{softmax}(\cdot)$ is the row-wise soft-max operation. The final output is formed by concatenation of the outputs of all heads, i.e,
\beq
{\bf X}_{\mathrm{out}}=\left[{\bf X}_1\ {\bf X}_2\ \ldots {\bf X}_{N_{\mathrm{head}}}\right] {\bf W}_{\mathrm{out}}\ \in \mathbb{R}^{N_q\times d_{\mathrm{model}}}
\eeq
where ${\bf W}_{\mathrm{out}}\ \in \mathbb{R}^{d_{\mathrm{model} \times d_{\mathrm{model}}}}$ is an additional learnable parameter.

For self-attention, when $N_q=N_{kv}$, further refinements such as residual connections and batch normalization $\mathrm{BN}(\cdot)$ (for stability while training) are appended. First we create a residual connection as
\beq
{\bf X}^\prime_{\mathrm{out}} =\mathrm{BN}\left({\bf X}_{\mathrm{v}} +{\bf X}_{\mathrm{out}}\right) \in \mathbb{R}^{N_q\times d_{\mathrm{model}}}
\eeq
and thereafter, non-linear activation is performed as 
\beq
{\bf X}^{\prime \prime}_{\mathrm{out}}=\mathrm{BN}\left(\mathrm{GELU}\left({\bf X}^\prime_{\mathrm{out}} {\bf W}_{o1} \right) {\bf W}_{o2} \right) \in \mathbb{R}^{N_q\times d_{\mathrm{model}}}
\eeq
where ${\bf W}_{o1}$ ($\in \mathbb{R}^{d_{\mathrm{model}} \times 4 d_{\mathrm{model}}}$) 
and ${\bf W}_{o2}$ ($\in \mathbb{R}^{4 d_{\mathrm{model}} \times d_{\mathrm{model}}}$) are additional learnable parameters. The Gaussian error linear unit activation is given by $\mathrm{GELU}(\cdot)$. The final output of the self-attention mechanism is ${\bf X}^{\prime \prime}_{\mathrm{out}}$.

\begin{figure}[htbp]
  \begin{minipage}{0.98\linewidth}
    \begin{center}
      \epsfig{figure=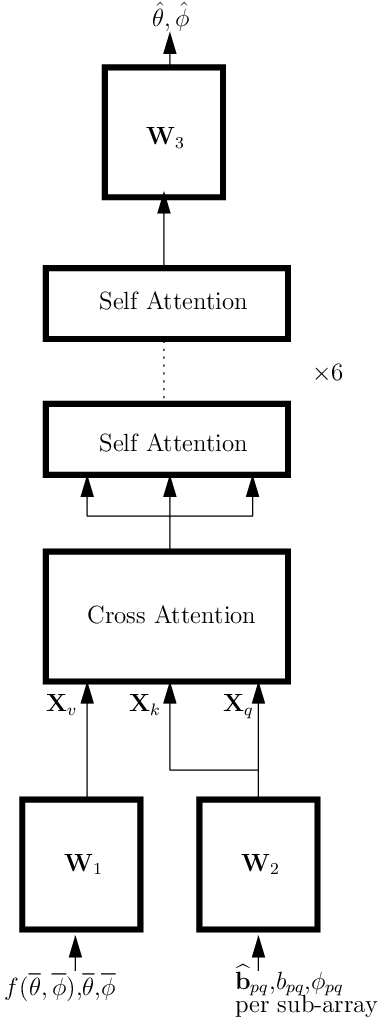,width=5.0cm}\\
    \end{center}
  \end{minipage}
  \caption{The transformer deep neural network architecture with input encoding (${\bf W}_1$,${\bf W}_2$) and output encoding (${\bf W}_3$), one cross attention block and 6 self attention blocks.\label{fig:transformer}}
\end{figure}

The full neural network architecture used in the proposed method is shown in Fig. \ref{fig:transformer}. We divide the input in terms of the aforementioned query, key and value triplet as follows:
\begin{itemize}
  \item The value: We compute the cost (\ref{fcost}) on a 2D grid $N_g\times N_g$ of $\theta$ ($\in [0,\pi/2]$) and $\phi$ ($\in [0,2\pi]$), $f(\overline{\theta},\overline{\phi})$ where we use $\overline{\theta}$ and $\overline{\phi}$ to denote the discretized values on the grid. Positional encoding is added as an additional input, where we also feed the discretized grid values $\overline{\theta}$ and $\overline{\phi}$ together with $f(\overline{\theta},\overline{\phi})$. Therefore, the input has 3 channels making the total size $3\times N_g \times N_g$.
  \item The query and the key: We consider to have $N_l$ sub-arrays with $N_r$ receivers in each sub-array. Let receivers $p$ and $q$ be the lowest indices in one such sub-array, and for this sub-array, we calculate the baseline unit vector $\widehat{\bf b}_{pq}$ and the baseline length $b_{pq}$ as well as the unwrapped phase $\phi_{pq}$ from (\ref{phi}). Therefore, for each sub-array, we have $5$ real values, making the total input size $5\times N_l$.
\end{itemize}
The original use of the transformer architecture is to process sequences of language tokens \citep{Vaswani}. In order to apply this to our problem, we need to create a sequence of patches from the 2D grid of $f(\overline{\theta},\overline{\phi})$. We divide the $N_g\times N_g$ grid into patches of size $N_p\times N_p$, yielding a sequence of $N_{kv}=(N_g/N_p)^2$ patches. The matrix ${\bf W}_1$ in Fig. \ref{fig:transformer} maps the $3\times N_p \times N_p$ patches into $d_{model}$ embedding dimension. In order to have $N_{kv}=N_q$, the matrix ${\bf W}_2$ in Fig. \ref{fig:transformer} maps the $5 N_l$ input into $d_{model}\times N_{kv}$. Therefore, we have $N_q=N_{kv}=(N_g/N_p)^2$.

The first transformer block uses cross attention where the value is different to the query and the key. Thereafter, 6 self attention blocks are cascaded. Finally, the output $\widehat{\theta}\in[0,\pi/2]$,$\widehat{\phi}\in[0,2\pi]$ is produced. The matrix ${\bf W}_3$ in Fig. \ref{fig:transformer} is mapping from $d_{model}$ to $2$ to produce the output. 

The trainable parameters in Fig. \ref{fig:transformer} are in ${\bf W}_1$,${\bf W}_2$, ${\bf W}_3$ and in the 7 attention blocks. They are trained by using simulated data where we provide the ground truth $\theta$ and $\phi$ to calculate $\widehat{\bf s}$ in (\ref{s}). The loss function is based on cosine similarity, i.e.,
\beq \label{loss}
\mathrm{loss}(\widehat{\theta},\widehat{\phi})=1-{\bf y}(\widehat{\theta},\widehat{\phi})^T \widehat{\bf s}
\eeq
where
\beq \label{loss1}
{\bf y}(\widehat{\theta},\widehat{\phi})\buildrel\triangle\over=\left[ \begin{array}{c} 
  \cos\widehat{\theta} \cos\widehat{\phi}\\
  \cos\widehat{\theta} \sin\widehat{\phi}\\  
\sin\widehat{\theta} \end{array} \right].
\eeq

For certain arrays, there is inherent ambiguity in the estimation of the DOA as we see in section \ref{sec:simul}. In such situations, we minimize (\ref{loss}) by using the minimum value of $\mathrm{loss}(\widehat{\theta},\widehat{\phi})$ and $\mathrm{loss}(\widehat{\theta},\widehat{\phi}+\pi)$, i.e.,
\beq \label{minloss}
\underline{\mathrm{loss}}(\widehat{\theta},\widehat{\phi})=\mathrm{min}\left(\mathrm{loss}(\widehat{\theta},\widehat{\phi}),\mathrm{loss}(\widehat{\theta},\widehat{\phi}+\pi)\right).
\eeq

\section{Simulations\label{sec:simul}}
\begin{figure}[htbp]
  \begin{minipage}{0.98\linewidth}
    \begin{center}
      \epsfig{figure=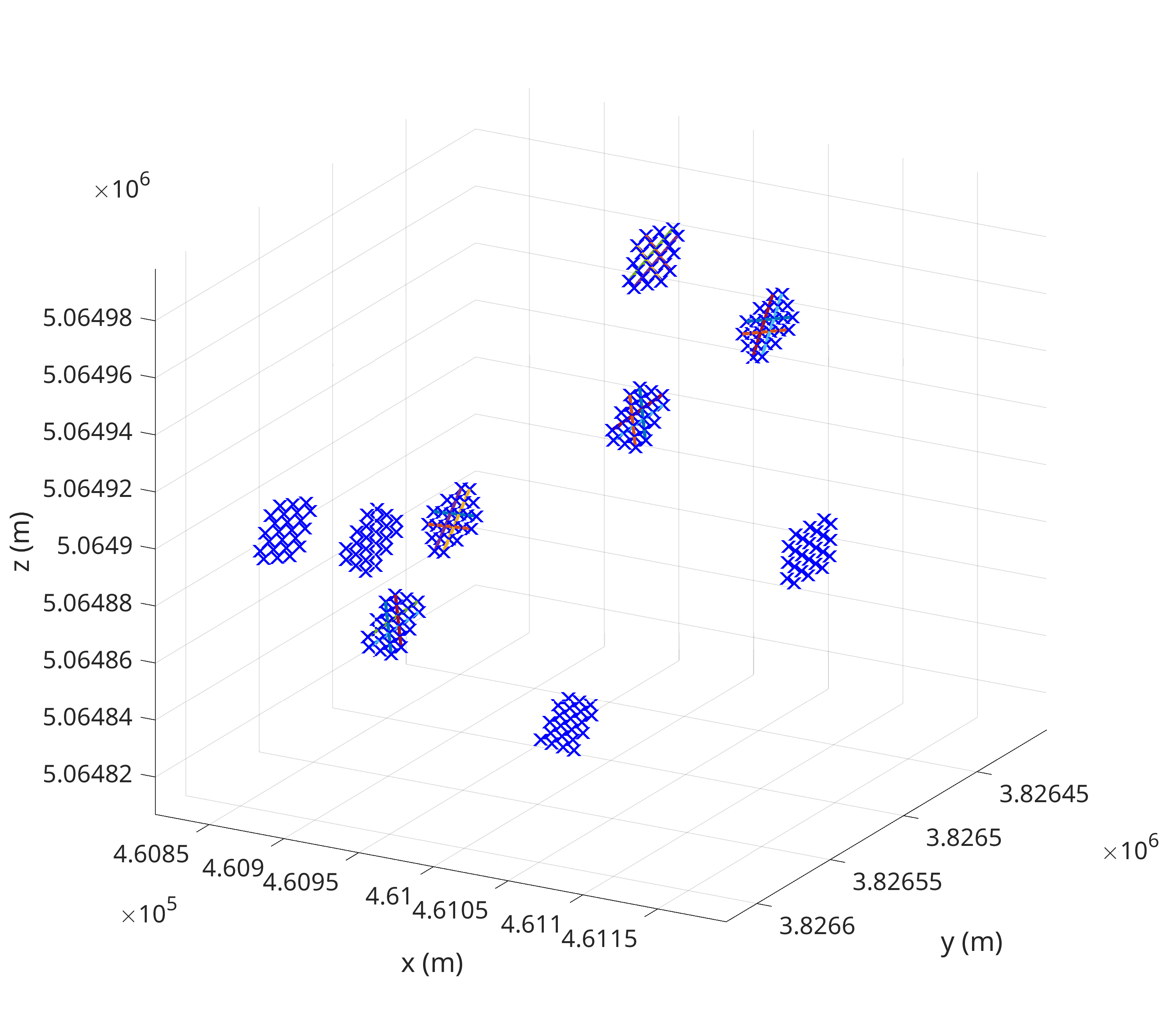,width=9.0cm}\\
    \end{center}
  \end{minipage}
  \caption{A fraction of the LOFAR AARTFAAC array receiver locations and the linear sub-arrays ($N_l$) are denoted by the various colours. Each sub-array has $N_r$ receivers.\label{fig:a12pos}}
\end{figure}

In this section, we provide simulations based on two existing radio interferometric array layouts, namely \citep[AARTFAAC]{AARTFAAC} and \citep[SKA-Low]{SKA0}. The first example is based on the AARTFAAC array, that consists of 576 LOFAR \citep{LOFAR} high band antenna receivers. These receivers are grouped and placed on regular grid points in 3D as seen in Fig. \ref{fig:a12pos}.

\begin{figure}[htbp]
  \begin{minipage}{0.98\linewidth}
    \begin{center}
      \epsfig{figure=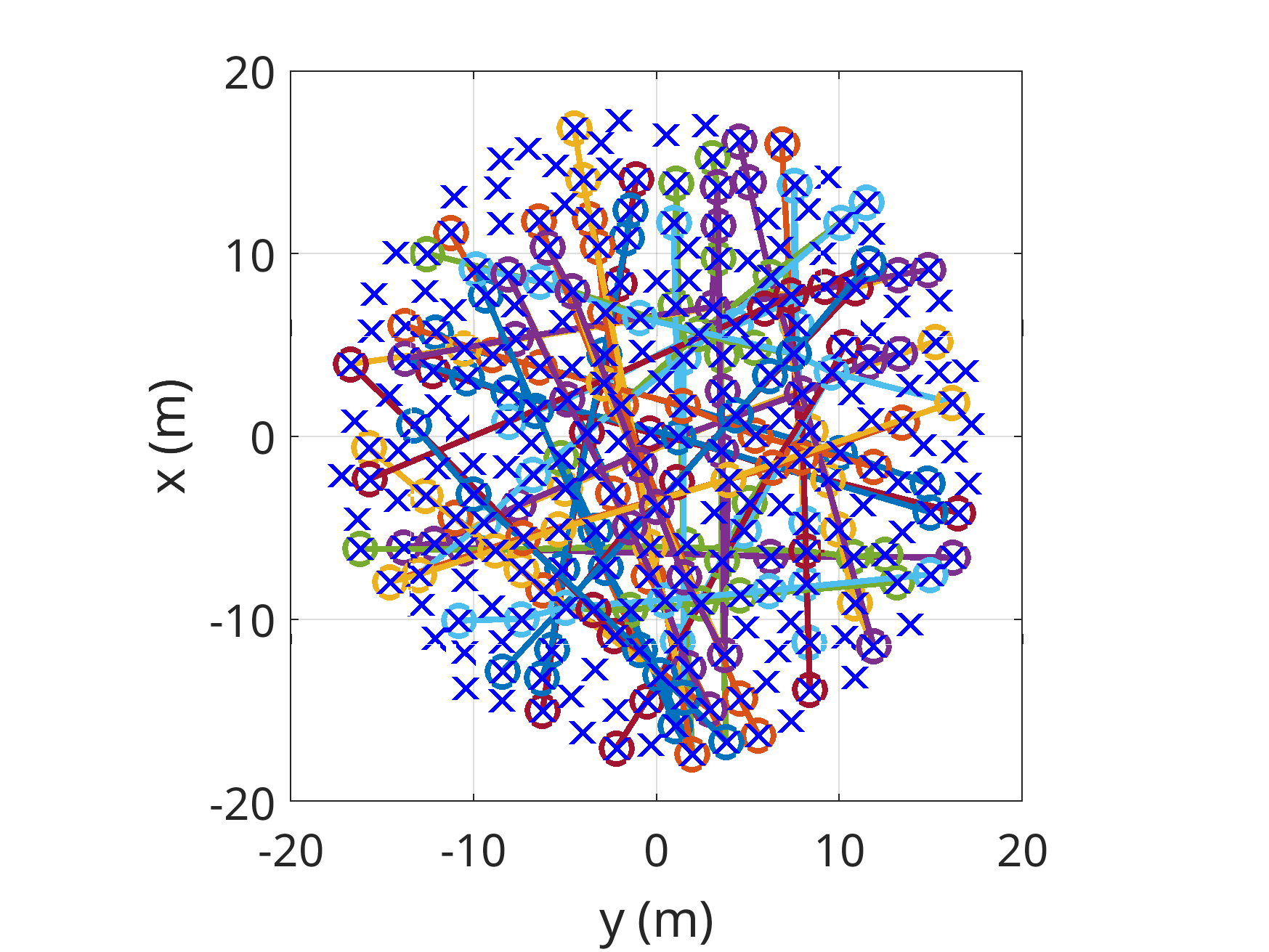,width=10.0cm}\\
    \end{center}
  \end{minipage}
  \caption{The layout of the receivers of a SKA-Low station. The receivers lie on a plane at random positions. Subsets of receivers are selected to form $N_l$ almost linear sub-arrays denoted by the various colours. Each sub-array has $N_r$ receivers.\label{fig:skapos}}
\end{figure}

We select the receivers of a single station of the SKA-Low \citep{SKA0} telescope as the second example. There are 256 dual-polarized receivers in this array, that are on a plane (hence 2D) as shown in Fig. \ref{fig:skapos}. Noteworthy in this layout is that the receivers are placed at random, hence almost linear sub-arrays are chosen by selecting subsets of receivers as shown in Fig. \ref{fig:skapos}.

From both these examples, we form $N_l$ sub-arrays that are (almost) linear by selecting $N_r$ receivers for each sub-array. In Figs. \ref{fig:a12pos} and \ref{fig:skapos}, the coloured lines denote the sub-arrays that are formed. Further details of the simulations and the deep learning models are given in Table \ref{configs}.
\begin{table}[!htbp]
\centering
  \caption{The simulation and deep learning settings for both example array configurations.\label{configs}}
\begin{tabular}{l|c|c|}
  {} & AARTFAAC & SKA\\\hline
  {\bf Configuration} & \multicolumn{2}{c|}{}\\
Total receivers &  576 & 256   \\
Selected receivers &  240 & 171   \\
Sub-arrays $N_l$  &  48 & 53   \\
Receivers per sub-array $N_r$  & \multicolumn{2}{c|}{6} \\\hline
  {\bf DOA simulation} & \multicolumn{2}{c|}{}\\
  DOA $\theta$ & \multicolumn{2}{c|}{$\mathcal{U}(0,\pi/2)$ rad} \\
  DOA $\phi$ & \multicolumn{2}{c|}{$\mathcal{U}(0,2\pi)$ rad} \\
  Range $R$ & \multicolumn{2}{c|}{$\mathcal{U}(100,100000)$ km} \\
  Polarization $\gamma$ & \multicolumn{2}{c|}{$\mathcal{U}(0,\pi/2)$ rad} \\
  Polarization $\eta$ & \multicolumn{2}{c|}{$\mathcal{U}(-\pi,\pi)$ rad} \\
  Power $E_s$& \multicolumn{2}{c|}{$\mathcal{U}(5,10)$ } \\
  Beam ${\bf E}_{p}$ $2\times2$ elements & \multicolumn{2}{c|}{$\mathcal{N}(0,1)+\jmath \mathcal{N}(0,1)$} \\
  Frequency $f$ & \multicolumn{2}{c|}{$\mathcal{U}(10,170)$ MHz} \\
  SNR & \multicolumn{2}{c|}{$\mathcal{U}(50,100)$} \\\hline
  {\bf Deep learning} &  \multicolumn{2}{c|}{}\\
  Grid size $N_g$&  \multicolumn{2}{c|}{128}\\
  Patches $N_p$&  \multicolumn{2}{c|}{16}\\
  Embedding dimension $d_{model}$  & 64 & 96\\
  Heads $N_{heads}$&  \multicolumn{2}{c|}{8}\\
Training samples &  \multicolumn{2}{c|}{60000}\\
Training iterations &  \multicolumn{2}{c|}{600000}\\
Testing samples &  \multicolumn{2}{c|}{10000}\\
  Learning rate &  \multicolumn{2}{c|}{$10^{-5}$}\\
  Loss  &  (\ref{loss}) & (\ref{minloss}) \\
Optimizer &  \multicolumn{2}{c|}{Adam}\\\hline
\end{tabular}
\end{table}

\begin{figure}[htbp]
  \begin{minipage}{0.98\linewidth}
    \begin{center}
      \epsfig{figure=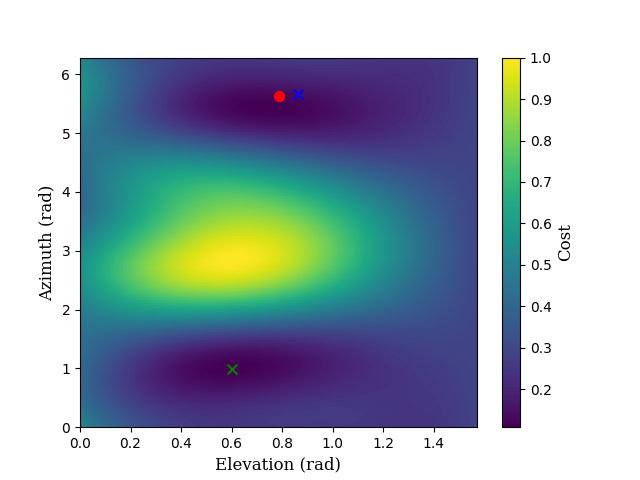,width=10.0cm}\\
    \end{center}
  \end{minipage}
  \caption{Cost function (\ref{fcost}) evaluated on a $N_g\times N_g$ grid of elevation $\theta$ and azimuth $\phi$ for the AARTFAAC array example. The ground truth DOA is shown by the red circle \textcolor{red}{o}. The DOA corresponding to the minimum value of the cost function is shown by the green cross \textcolor{green}{x}. The output of the transformer deep neural network is shown by the blue cross \textcolor{blue}{x}.\label{fig:a12_eval}}
\end{figure}

\begin{figure}[htbp]
  \begin{minipage}{0.98\linewidth}
    \begin{center}
      \epsfig{figure=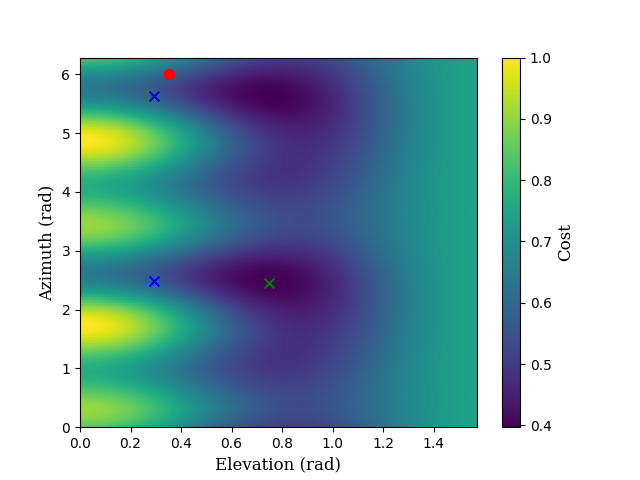,width=10.0cm}\\
    \end{center}
  \end{minipage}
  \caption{Cost function (\ref{fcost}) evaluated on a $N_g\times N_g$ grid of elevation $\theta$ and azimuth $\phi$ for the SKA-Low array example. Noteworthy is the symmetry of the cost with the azimuth $\phi$. This symmetry is due to the array having a 2D layout yielding an ambiguity in $\phi$. The ground truth DOA is shown by the red circle \textcolor{red}{o}. The DOA corresponding to the minimum value of the cost function is shown by the green cross \textcolor{green}{x}. The output of the transformer deep neural network (taking into account the ambiguity of $\pi$) is shown by the blue crosses \textcolor{blue}{x}.\label{fig:ska_eval}}
\end{figure}

The data for training the deep learning models are generated separately for each array using the settings in Table \ref{configs} that implement (\ref{voltage}). For 30\% of the simulations, the value of $R$ is chosen to be finite (near-field) and for other cases, we simulate a far-field source. Thereafter, the phases $\psi_{pq}$ in (\ref{wrapped_psi}) are estimated for baselines that belong to each of the $N_l$ sub-arrays using ESPRIT. Next, the phase of each sub-array is unwrapped using the phase difference projection algorithm. Finally, the cost function (\ref{fcost}) is evaluated on a grid of $N_g\times N_g$. In Figs. \ref{fig:a12_eval} and \ref{fig:ska_eval}, we show two examples of the evaluated cost functions for the two arrays. We notice the symmetry of the cost function with the azimuth $\phi$ for the SKA-Low array configuration in Fig. \ref{fig:ska_eval}. This is due to the 2D (planar) layout of the receiver locations seen in Fig. \ref{fig:skapos}.

During training the deep neural network, the discretized cost function $f(\overline{\theta},\overline{\phi})$ and assorted metadata are fed as input to the model in Fig. \ref{fig:transformer} together with the ground truth DOA, i.e., $\theta$ and $\phi$. We use the Adam optimizer \citep{Adam} for training the model using the settings listed in Table \ref{configs}. Note that for the SKA-Low array example, the cost function for training is modified as (\ref{minloss}) to take into account the inherent ambiguity.

After training, we generate an additional $10000$ samples of data using the same settings as in Table \ref{configs} for testing the model accuracy (30\% near-field sources). In Figs. \ref{fig:a12_eval} and \ref{fig:ska_eval}, we have shown the DOA estimates given by the trained DNN as well (together with the ground truth). For quantitative evaluation, we calculate the angular error between the estimated DOA $\widehat{\theta}$,$\widehat{\phi}$ (direction vector ${\bf y}(\widehat{\theta},\widehat{\phi})$ (\ref{loss1})) and the ground truth DOA $\theta$,$\phi$ (direction vector $\widehat{\bf s}$ (\ref{s})) as $\mathrm{arccos}({\bf y}^T \widehat{\bf s})$ for all $10000$ samples.

\begin{figure}[htbp]
  \begin{minipage}{0.98\linewidth}
    \begin{center}
      \epsfig{figure=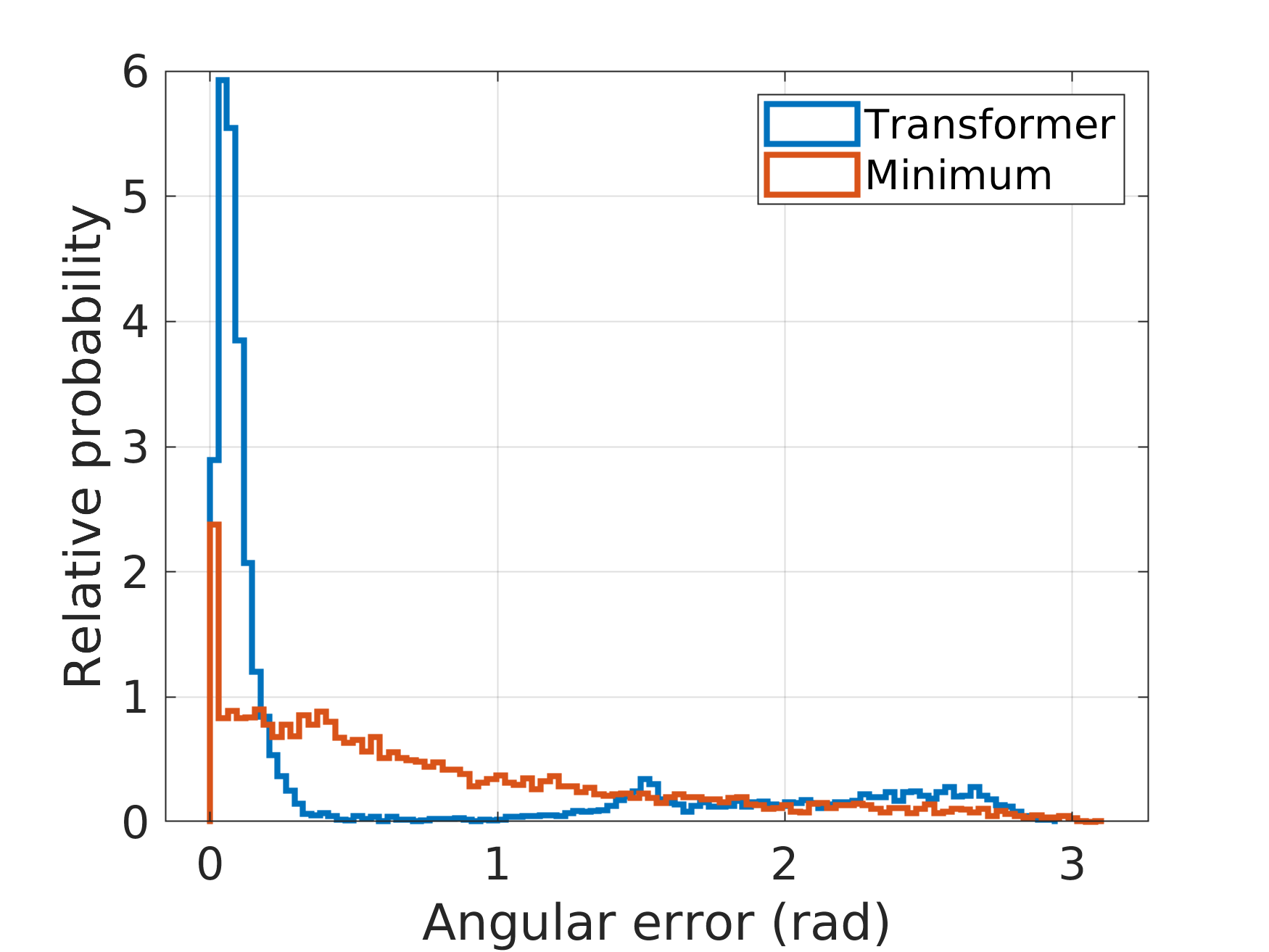,width=9.0cm}\\
    \end{center}
  \end{minipage}
  \caption{The histogram of the angular error for the AARTFAAC array configuration using 10000 samples. The angular error for DOA estimation using the DNN as well as for DOA estimation using the minimum value of (\ref{fcost}) are shown. The histogram peaks around 3 deg for the DOA estimation using the DNN.\label{fig:a12_error}}
\end{figure}

\begin{figure}[ht]
  \begin{minipage}{0.98\linewidth}
    \begin{center}
      \epsfig{figure=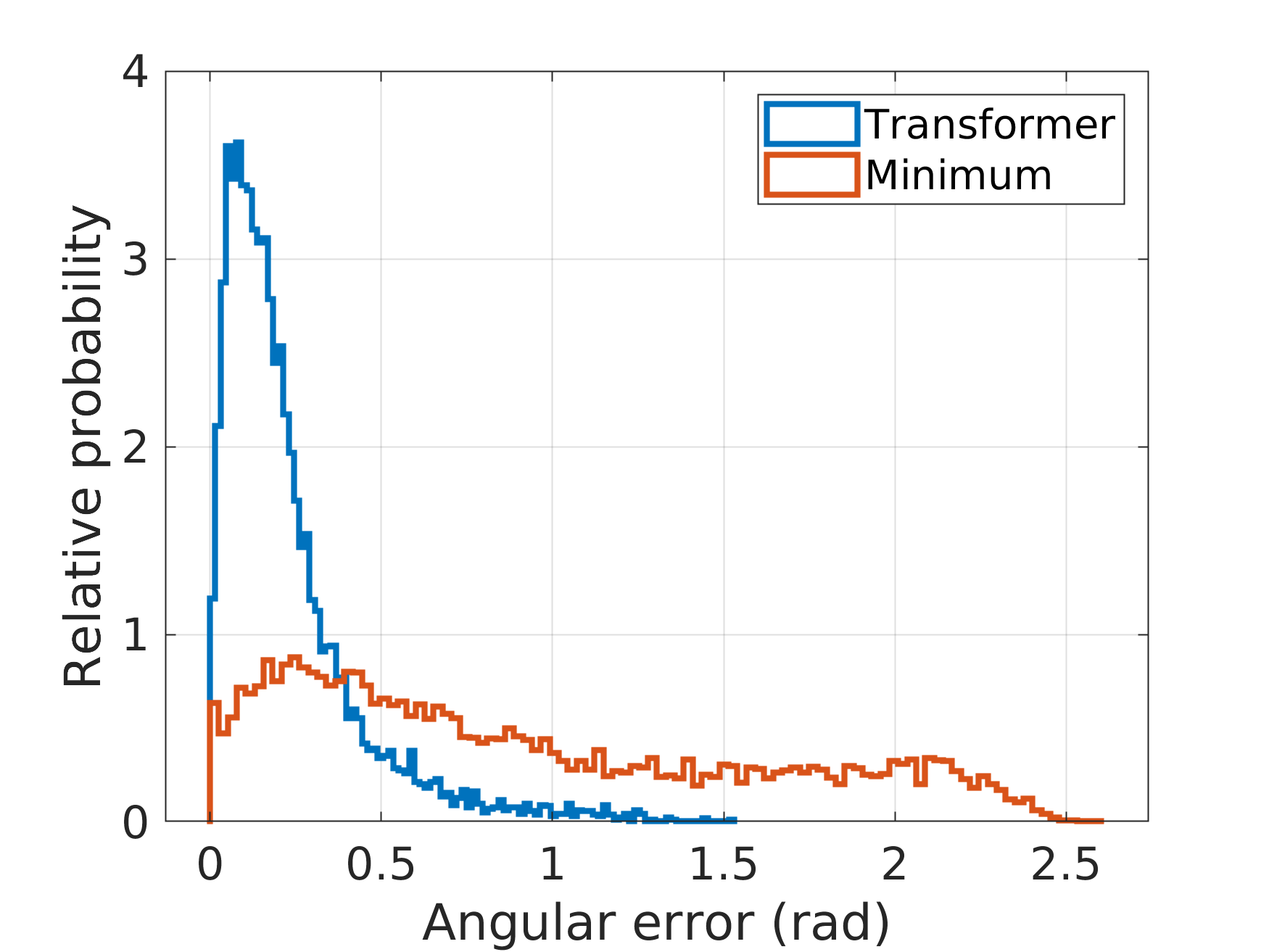,width=9.0cm}\\
    \end{center}
  \end{minipage}
  \caption{The histogram of the angular error for the SKA-Low array configuration using 10000 samples. The angular error for DOA estimation using the DNN as well as for DOA estimation using the minimum value of (\ref{fcost}) are shown. The histogram peaks around 5 deg for the DOA estimation using the DNN. The spread of the angular error is broader than the result in Fig. \ref{fig:a12_error}. We attribute that to receiver locations not entirely being co-linear, hence the sub-arrays that are chosen are almost but not entirely linear.\label{fig:ska_error}}
\end{figure}

The histogram of angular error for the DOA estimation using the trained DNN and using the minimum value of (\ref{fcost}) for both arrays are shown in Figs. \ref{fig:a12_error} and \ref{fig:ska_error}. Note that because of the inherent ambiguity, the angular error for SKA-Low array is calculated by considering both $\widehat{\phi}$ and $\widehat{\phi}+\pi$ as the possible estimates for the azimuth angle. Also noteworthy is the bias of the phase estimate (the peak of the histogram is not at zero), which is also reported in previous work \citep{Bell2025}.

\begin{figure}[htbp]
  \begin{minipage}{0.98\linewidth}
    \begin{center}
      \epsfig{figure=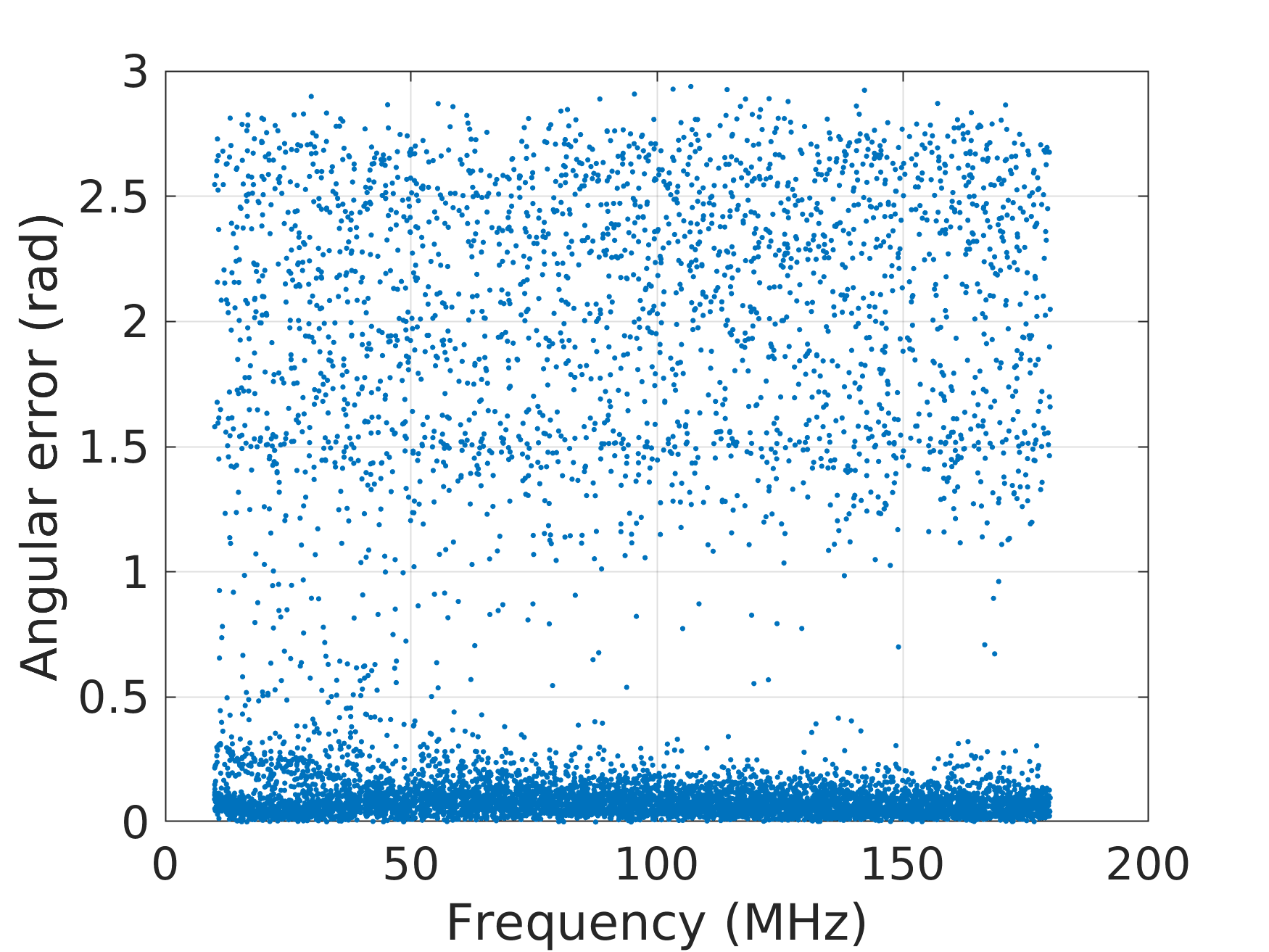,width=9.0cm}\\
    \end{center}
  \end{minipage}
  \caption{The angular error for the AARTFAAC array configuration using the DNN output versus the frequency $f$ of the simulated data for the 10000 samples. At each frequency, the effective array spacing will change and hence the phase unwrapping needs to pre-calculate the projection parameters. We see almost uniform error spread across frequency.\label{fig:a12_error_freq}}
\end{figure}

\begin{figure}[ht]
  \begin{minipage}{0.98\linewidth}
    \begin{center}
      \epsfig{figure=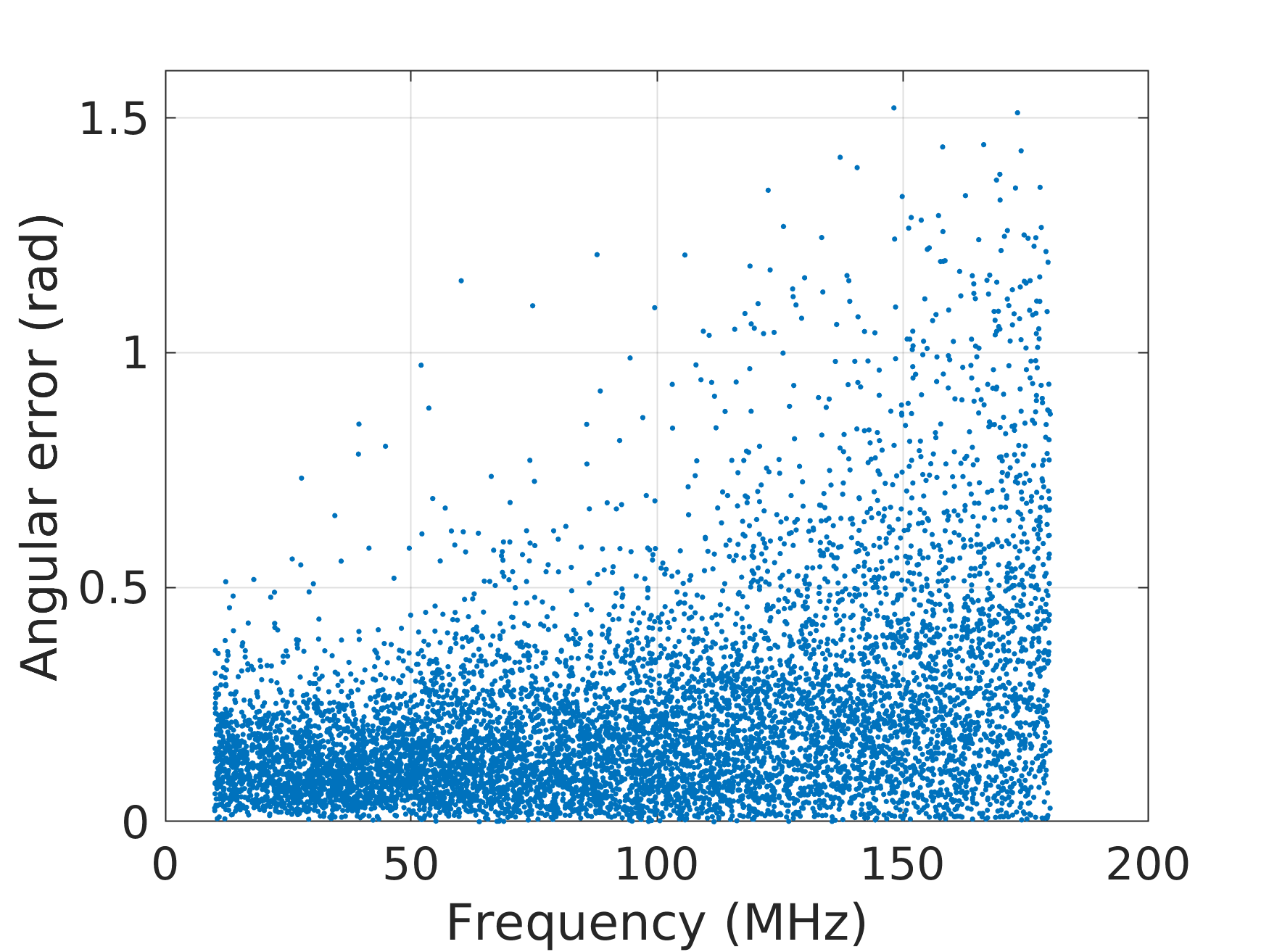,width=9.0cm}\\
    \end{center}
  \end{minipage}
  \caption{The angular error for the SKA-Low array configuration using the DNN output versus the frequency $f$ of the simulated data for the 10000 samples. We see a broader spread of the angular error compared to Fig. \ref{fig:a12_error_freq}. Note also that the upper bound of error is reduced to $\pi/2$ because we consider the ambiguity in $\phi$ when calculating the error.\label{fig:ska_error_freq}}
\end{figure}

In order to evaluate the wideband performance of the DOA estimation, we plot the angular error variation with frequency range in the simulation, i.e., $[10,170]$ MHz in Figs. \ref{fig:a12_error_freq} and \ref{fig:ska_error_freq}. Note that the actual operating frequency ranges of both these arrays are different than what is simulated, but nonetheless it is much wider than the operating frequency of a typical array dedicated for DOA estimation. We see from Figs. \ref{fig:a12_error_freq} and \ref{fig:ska_error_freq} that we get almost uniform performance over the frequency range. The crucial aspect for this performance is the phase unwrapping, where the PDP algorithm needs to perform some pre-calculations per each frequency.

We discuss the computational cost of the proposed method with respect to existing methods briefly. We consider the computational cost of the full eigenvalue decomposition of an $N\times N$ matrix to be $O(N^3)$. We need to find the dominant eigenvector by performing the eigenvalue decomposition as in (\ref{eig}) where $N=4$, hence the complexity is $O(4^3)$ (Note that we can use methods like the power iteration that have lower complexity, but we keep this larger complexity here). Therefore, the total complexity of running ESPRIT is $N_l \frac{N_r (N_r-1)}{2} O(4^3)$ for the full set of sub-arrays. The other steps such as phase unwrapping and evaluating the DNN have fixed and lower computational complexity. In contrast, for running an algorithm like MUSIC \citep{MUSIC}, considering the full array to have $N_l N_r$ receivers, the complexity of the eigenvalue decomposition of the full correlation matrix is $O((N_l N_r)^3)$ and the search in $N_g^2$ grid points involve matrix-vector multiplication with complexity $O((N_l N_r)^2 N_g^2)$. Sparsity based methods \citep{YANG2018,Yimin2024} need to create the dictionary matrix with $N_g^2$ or more steering vectors in 2D and will have matrices of size $(N_l N_r)^2 \times N_g^4$ (because of the kronecker product). Hence, compared to both these popular methods, the proposed method is much less computationally demanding.

\begin{figure}[htbp]
  \begin{minipage}{0.98\linewidth}
\begin{center}
  \begin{minipage}{0.98\linewidth}
\centering
  \centerline{\includegraphics[width=0.8\textwidth]{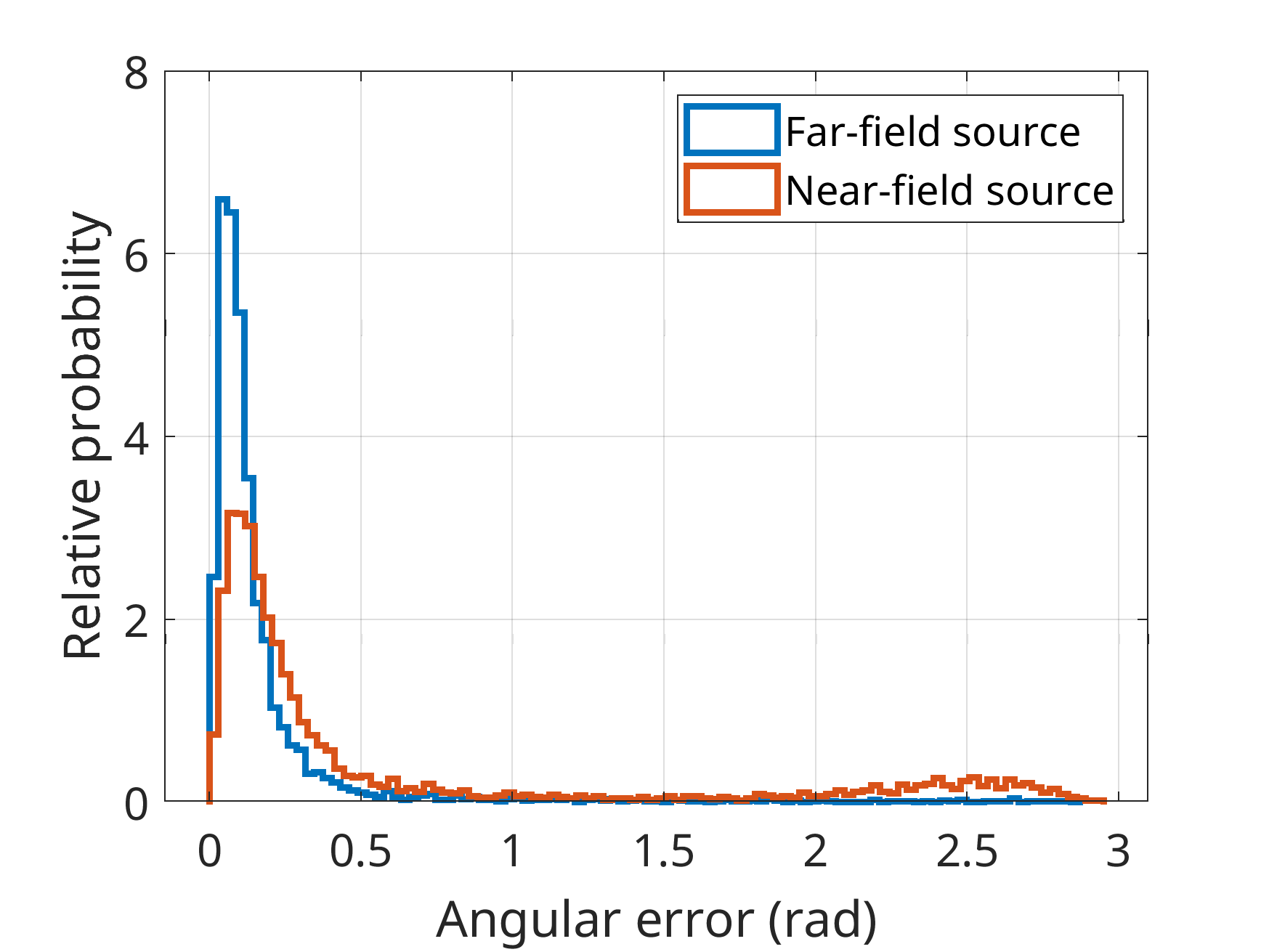}}
\vspace{0.1cm} \centerline{(a)}\smallskip
\end{minipage}
  \begin{minipage}{0.98\linewidth}
\centering
  \centerline{\includegraphics[width=0.8\textwidth]{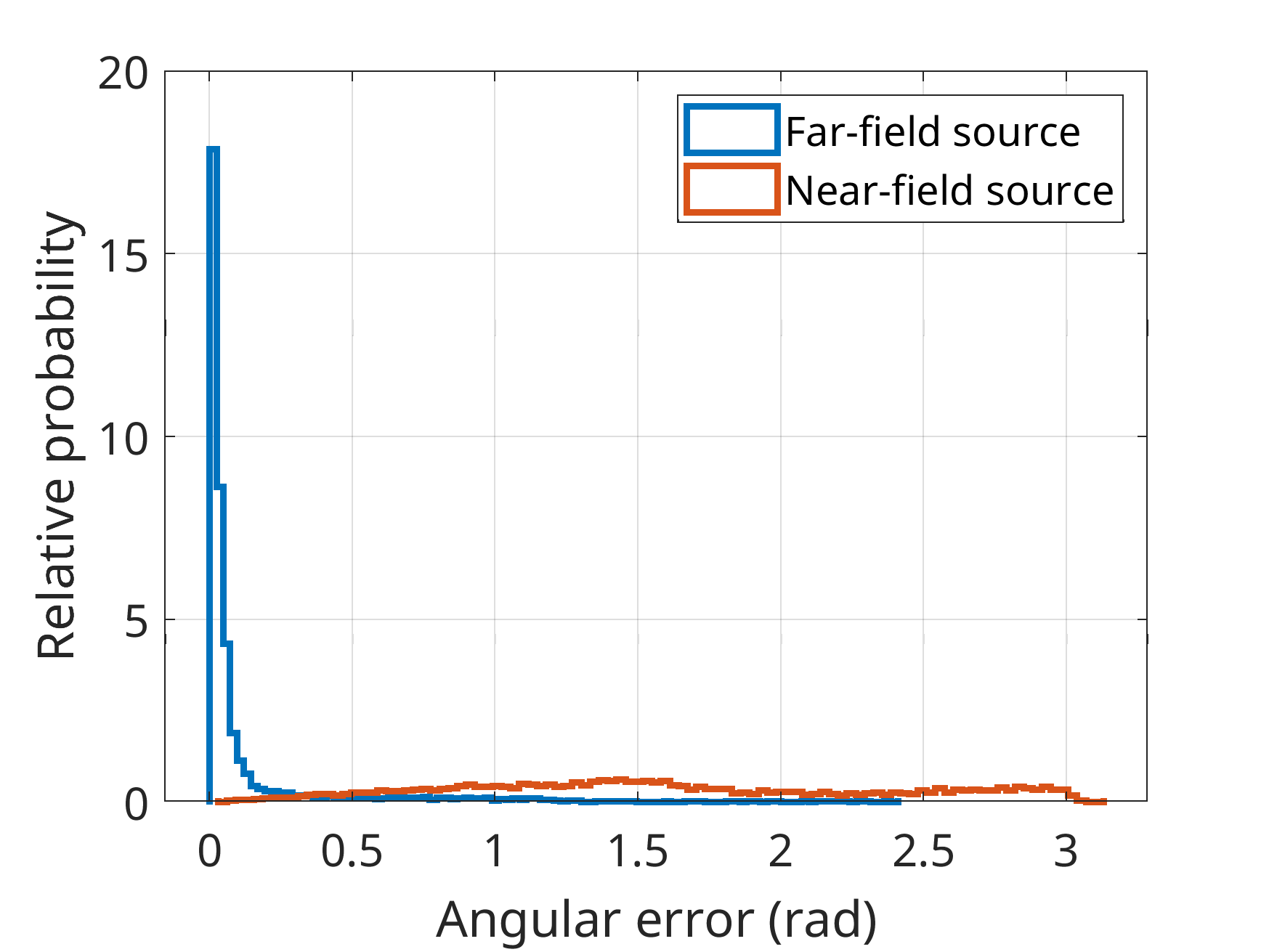}}
\vspace{0.1cm} \centerline{(b)}\smallskip
\end{minipage}
\end{center}
  \end{minipage}
  \caption{The comparison of the angular error of the transformer with MUSIC for 10000 samples. The histograms of the angular error for (a) the transformer DNN and (b) MUSIC \citep{MUSIC} are shown. In each panel we show the angular error for sources in the far-field as well as in the near-field. \label{fig:music_comparison}}
\end{figure}

A quantitative comparison of the proposed method (AARTFAAC array) with MUSIC is presented in Fig. \ref{fig:music_comparison}. We show the results for both far-field and near-field sources in this figure. For near-field sources, the MUSIC algorithm (using the full correlation matrix) gives better accuracy. The MUSIC algorithm is using steering vectors that are designed for a far-field source. Hence, as seen in Fig. \ref{fig:music_comparison} (b), when we have a source in the near-field of the array, the MUSIC algorithm designed for a far-field source completely fails to perform. In such a situation, both a direction and range search should be performed separately in MUSIC and the separation of the direction from the range is only possible for specialized array geometries and not for arbitrary arrays. In contrast, the proposed method gives accurate results both for far-field and near-field sources, without having to change the data model that is being used.

The computational time for the comparison shown in Fig. \ref{fig:music_comparison} was measured using a CPU with 8 threads. We measure the average time for one realization in the following results. The data used for the comparison were generated with 1 s duration and 1000 samples to form the correlation matrix. We consider the AARTFAAC array with $N_r=6$ and $N_l=48$ from Table \ref{configs}.
For the MUSIC algorithm, the time taken was 1.4 s (using 8 threads with the matrix size $2 N_l N_r \times 2 N_l N_r$). For the proposed method, the individual time measurements were as follows. For running ESPRIT on each sub-array (sequential, single thread, for $N_r(N_r-1)/2$ baselines) about 3e-4 s, consuming a total ($\times N_l$) time of 0.014 s. For phase unwrapping for each sub-array it consumed about 3e-4 s giving a total ($\times N_l$) time of 0.016 s. The evaluation of the DNN (transformer) took 0.004 s on the CPU. Therefore, the proposed algorithm took a total time of 0.034 s, which is much lower than the time taken by MUSIC.

\section{Conclusions\label{sec:conc}}
We have proposed a DOA estimation method that can be adopted by existing radio interferometric arrays with dual polarized receivers. The computational cost is reduced by performing per-baseline ESPRIT as the computationally demanding task. Further post-processing steps are needed such as phase unwrapping and the use of a transformer deep neural network. Simulations show that the proposed method provides satisfactory performance, especially in finding sources of RFI although only one dominant source of RFI is assumed. Future work will expand this method to handle multiple DOA estimation and the use of the proposed method in applications beyond radio astronomy.

Source code implementing all algorithms discussed in this paper are publicly accessible at (\href{https://github.com/SarodYatawatta/doaFind}{doaFind}).
\section*{Acknowledgments}
We dedicate this work to our passed colleague, Albert-Jan Boonstra. We also thank the anonymous reviewers for the careful review and valuable comments.
\appendix
\section{Range and DOA estimation for near-field sources\label{appendixA}}
For a near-field source, the source direction changes with respect to each baseline ${\bf b}_{pq}$. Therefore, we consider the centroid of each baseline to find the source direction vector $\widehat{\bf s}_{pq}$ as
\beq
\widehat{\bf s}_{pq}=\frac{R\widehat{\bf s}-({\bf x}_p+{\bf x}_q)/2}{\|R\widehat{\bf s}-({\bf x}_p+{\bf x}_q)/2\|}
\eeq
and we formulate a cost function that also includes the range $R$ in addition to the DOA as
\beq \label{fcost_nearfield}
f(R,\theta,\phi)=\sum_{pq} \left( \left(\widehat{\bf b}^T_{pq}\ \widehat{\bf s}_{pq} \right)^2 - \left(\frac{\phi_{pq}}{2 \pi b_{pq}/\lambda}\right)^2 \right)^2. 
\eeq

\begin{figure}[htbp]
  \begin{minipage}{0.98\linewidth}
\begin{center}
  \begin{minipage}{0.98\linewidth}
\centering
  \centerline{\includegraphics[width=0.8\textwidth]{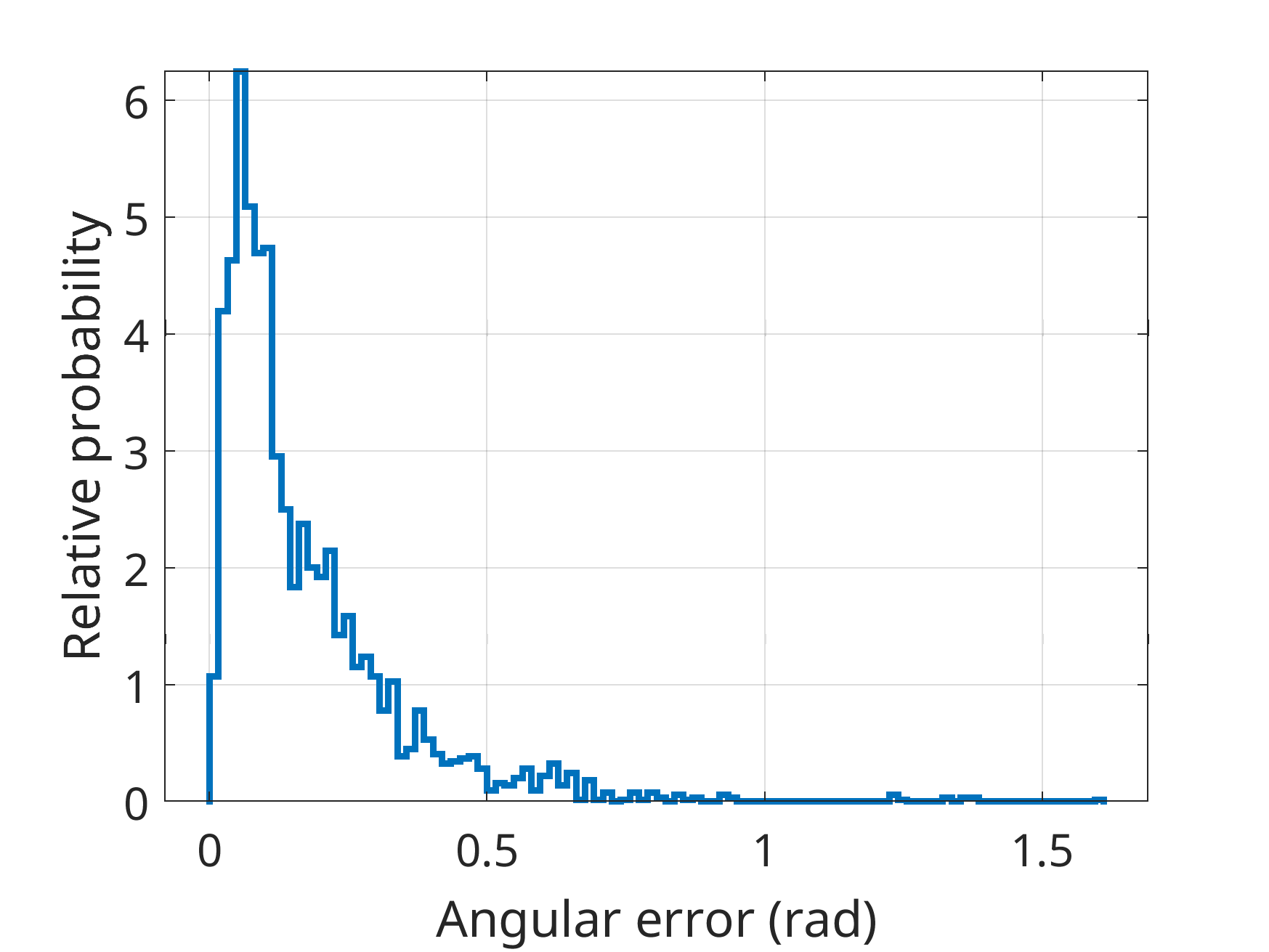}}
\vspace{0.1cm} \centerline{(a)}\smallskip
\end{minipage}
  \begin{minipage}{0.98\linewidth}
\centering
  \centerline{\includegraphics[width=0.8\textwidth]{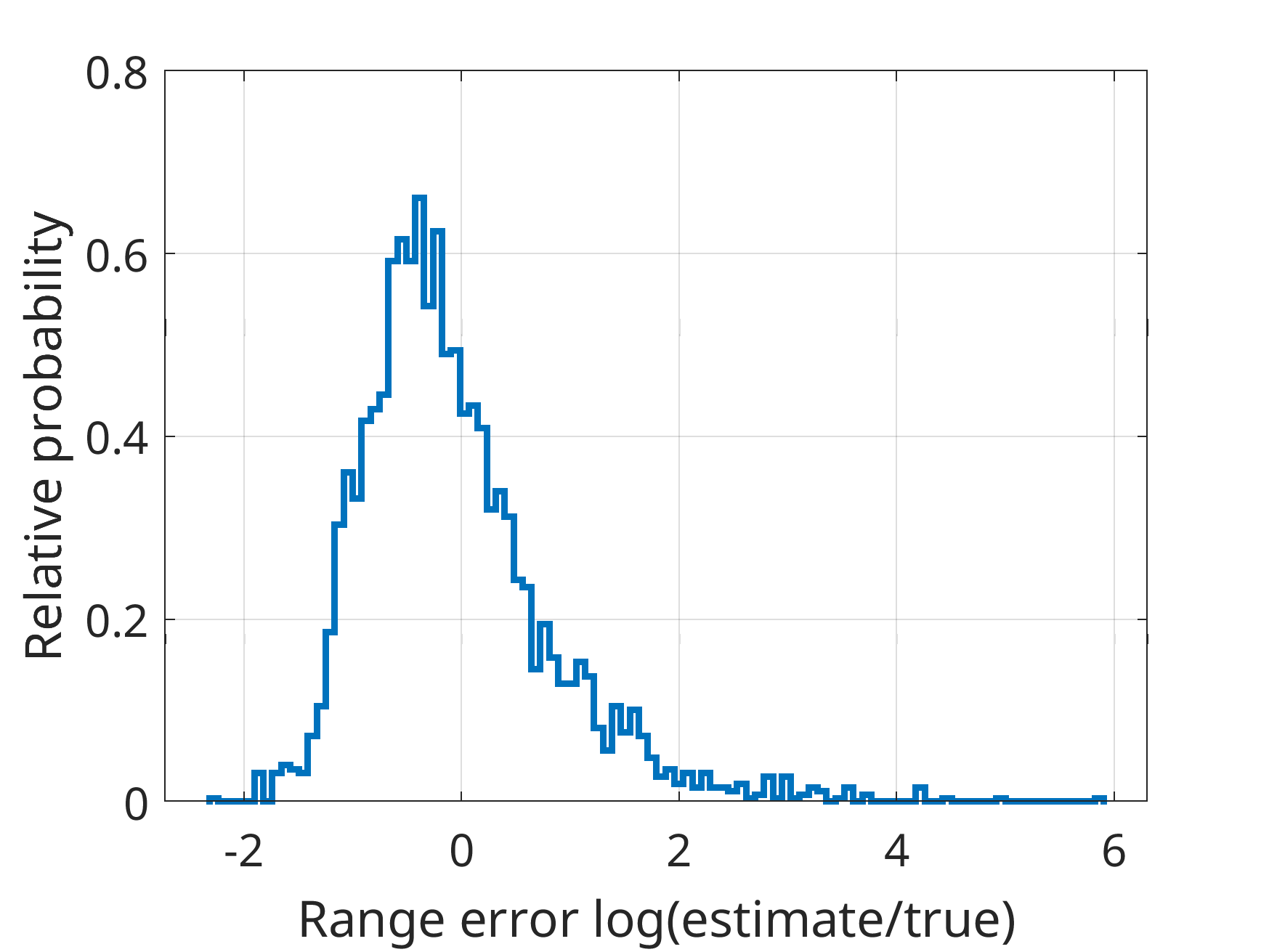}}
\vspace{0.1cm} \centerline{(b)}\smallskip
\end{minipage}
\end{center}
  \end{minipage}
  \caption{An illustration of the DOA and range estimation error of the transformer for the AARTFAAC array. The histograms of the (a) angular error and (b) range error are shown for 4000 examples of near-field sources. \label{fig:range_doa}}
\end{figure}

This function is evaluated on a grid of DOA $\overline{\theta}$,$\overline{\phi}$ as well as on a discrete set of range values, $\overline{R}$ as $f(\overline{R},\overline{\theta},\overline{\phi})$. The transformer DNN in Fig. \ref{fig:transformer} is modified to not only output the DOA $\hat{\theta}$,$\hat{\phi}$ but also the range $\log \hat{R}$. No additional input is fed to the DNN compared with the near-field case. In Fig. \ref{fig:range_doa}, we show an example where we train the transformer to not only estimate the DOA but also the range (AARTFAAC array). However, we only evaluate the range (log value) on 10 grid points uniformly sampled within the range given in Table \ref{configs}, so this is a very coarse estimate. Nevertheless, we are able to get range results with reasonable accuracy albeit with a bias (range underestimated).

\bibliographystyle{model2-names} 
\bibliography{references}





\end{document}